\documentclass[a4paper]{article}
\usepackage[11pt]{extsizes}
\usepackage[pdftex,colorlinks=true,linkcolor=blue,citecolor=blue,urlcolor=blue]{hyperref}
\usepackage{graphicx}
\usepackage{amsmath}
\usepackage{lscape}
\usepackage{geometry}
\usepackage{amsfonts}
\usepackage{amssymb}

\begin{document}

\title{\textbf{Saturation of a large scale instability and non linear stuctures in a rotating stratified flow} }
\author{Anatoly TUR, Malik CHABANE \\
Universit\'{e} de Toulouse [UPS],\\
CNRS, Institut de recherche en astrophysique et plan\'{e}tologie,\\
9 avenue du Colonel Roche, BP 44346, \\
31028 Toulouse Cedex 4, France \and Vladimir YANOVSKY \\
Institute for Single Crystals, National Academy of\\
Science Ukraine, Kharkov 31001, Ukraine}

\maketitle

\section*{Abstract}

\begin{abstract}

In this letter, we present non linear structures resulting from the saturation of a large scale instability in rotating stratified fluids with small scale forced turbulence which we found in \cite{[1]}. These structures are of helical kinks type. The different solutions that we plotted correspond to solutions linking two different stationnary points for different values of the Rayleigh and the Taylor numbers.
\bigskip

\bigskip Keywords: Large scale instability, Coriolis force, buoyancy, instability saturation, non linear stuctures
\end{abstract}

\section{Introduction}

Large scale instabilities are of a great interest in fluid dynamics. They generate vortices which play a
fundamental role in turbulence and in transport processes. The generation of this kind of instability has been widely studied, for example in \cite{[15]}, \cite{[18]}, \cite{[21]}, \cite{[26]}. Direct numerical simulations of Boussinesq Equation confirmed the existence of large scale vortex generation in stratified and rotating flows \cite{[14b]}. Many papers as well as the results of numerical modelling are described in detail in review \cite{[21]}. Several works have shown that a necessary condition for these instabilities to appear is the lack of reflection invariance or parity invariance \cite{[15]}, \cite{[16]}, \cite{[18]}. In fluid dynamics, the widespread mechanism used to break this symmetry is the helicity $\vec{v}\cdot \vec{rot}\vec{v}\neq 0$. However, it has been shown that the helicity itself is not enough to generate these large scale stucture, and others factors are necessary like for example temperature gradient \cite{[18]}, \cite{[26]}. It has also been shown in a previous work \cite{[26]} that the method of multi-scale development can lead to the occurence of large scale instabilities in helical stratified turbulence.\\
In the work \cite{[1]}, we found a new large scale instability in rotating stratified fluids with small scale forced turbulence. The helicity can be used in an explicit way \cite{[26]} as well as in a internal way \cite{[27]}. In this previous paper, we used the Coriolis force and the buyoancy to create naturally the internal helicity which is required to break the symmetry of the flow. That is what allowed us to use a force which does not have any particular properties (especially it
is nonhelical and it does not lack parity invariance). The force only
maintains turbulent fluctuations.\newline
In this letter, we complete this previous study by solving the non linear instabilty equations. As a result of saturation, the non linear structures appearing are rotational kinks.
Our letter is organized as follows: In section $2$ we remind the basics equations of the problem, in section $3$ we briefly remind the scheme of asymptotic development and write down the main equations for large scale instability which was found in \cite{[1]}. In section $4$, we expose the expression of the external force and the Reynolds stresses calculated in \cite{[1]}, in section $5$ we discuss the instability conditions. Section $6$ is devoted to the calculation of the non linear saturated instability, a brief discussion of the result is given as a conclusion in section $7$.

\bigskip

\section{The main equations and formulation of the problem}

Let's consider the equations for the motion of an incompressible fluid with
a constant temperature gradient in the Boussinesq approximation:

\begin{equation}
\frac{\partial \vec{V}}{\partial t}+(\vec{V}\cdot\nabla)\vec{V}+2\vec{\Omega} \times \vec{V}=-\frac{1}{\rho _{0}}\nabla P+\nu \Delta \vec{V}+g\beta T\vec{l}+\vec{F}_{0}  \label{1} \newline
\end{equation}

\begin{equation}
\frac{\partial T}{\partial t}+(\vec{V}\cdot\nabla )T=\chi \Delta T-V_{z}A.  \label{2}
\end{equation}

\begin{equation}
\nabla \cdot\vec{V}=0
\end{equation}

Where $\vec{l}=\left( 0,0,1\right) $, $\beta $ is the thermal expansion
coefficient, $A=\frac{dT_{0}}{dz}$ is the constant equilibrium gradient of
the temperature, $\rho _{0}=Const.$, and $\nabla T_{0}=A\vec{l}$.\\
Using dimensionless variables, we can rewrite the previous equation as:

\begin{equation}
\frac{\partial \vec{V}}{\partial t}+R(\vec{V}\cdot\nabla )
\vec{V}-\Delta \vec{V}+D\vec{l}\times\vec{V}=-\nabla P+RaT\vec{l}+\vec{F}_{0}  \label{3}
\end{equation}

\begin{equation}
\left( \frac{\partial T}{\partial t}-\Delta T\right) =-V_{z}-R(\vec{V}\cdot\nabla )T  \label{4}
\end{equation}

\begin{equation*}
\nabla\cdot\vec{V}=0
\end{equation*}

\noindent where $R$ and $D$ are respectively the Reynolds number and the scare root of the Taylor numbers. $Pr$ represents the Prandtl number and $Ra$ is the Rayleigh number. We also introduce a dimensionless temperature $T$.

We will consider as a small parameter of an asymptotic development the
Reynolds number $R$. Let us denote the small scale variables by $x_{0}=(\vec{x}%
_{0},t_{0})$ , and the large scale ones by $X=(\vec{X},T)$. The small scale
partial derivative operation $\frac{\partial }{\partial x_{0}^{i}},\frac{%
\partial }{ \partial t_{0}}$, and the large scale ones $\frac{\partial }{%
\partial \vec{X}}, \frac{\partial }{\partial T}$ are written, respectively,
as $\partial _{i},\partial _{t},\nabla _{i}$ and $\partial _{T}$. To
construct a multi-scale asymptotic development we follow the method which is
proposed in \cite{[15]}.

\section{The multi-scale asymptotic development}

Following \cite{[1]}, let us look for the solutions to Equations (\ref{3}) and (\ref{4}) in the
following form: 

\begin{equation}
\vec{V}(\vec{x},t)=\frac{1}{R}\vec{W}
_{-1}(X)+\vec{v}_{0}(x_{0})+R\vec{v}_{1}+R^{2}\vec{v}_{2}+R^{3}\vec{v}_{3}+\cdots  \label{68.1}
\end{equation}

\begin{equation}
T(\vec{x},t)=\frac{1}{R}T_{-1}(X)+T_{0}(x_{0})+RT_{1}+R^{2}T_{2}+R^{3}T_{3}+\cdots  \label{68.2}
\end{equation}

\begin{equation}
P(\vec{x},t)=\frac{1}{R^{3}}P_{-3}(X)+\frac{1}{R^{2}}P_{-2}(X)+\frac{1}{R}P_{-1}(X)+P_{0}(x_{0})+R(P_{1}+\overline{P}_{1}(X))+R^{2}P_{2}+R^{3}P_{3}+\cdots
\end{equation}

We remind that the scale relation is the following: $\vec{X}=R^2\vec{x}_{0}$ and $%
T=R^4t_{0}$.

From this we get the main secular equation at order $\cal{O}$$(R^3)$:

\begin{equation}
\partial _{T}W_{-1}^{i}-\Delta W_{-1}^{i}+\nabla _{k}(\overline{%
v_{0}^{k}v_{0}^{i}})=-\nabla _{i}\overline{P}_{1},  \label{68.4}
\end{equation}

\begin{equation}
\partial _{T}T_{-1}-\Delta T_{-1}+\nabla _{k}(\overline{v_{0}^{k}T_{0}})=0.
\label{68.5}
\end{equation}

\section{Calculations of the Reynolds stresses}

The essential equation for finding the nonlinear
alpha-effect is equation (\ref{68.4}). In order to obtain these equations in
closed form, we need to calculate the Reynolds stresses $\nabla _{k}(%
\overline{v_{0}^{k}v_{0}^{i}})$.\\
The external force can be chosen in a general $3D$ form like for example:

\begin{equation}
\vec{F}_{0}=f_{0}\left( \vec{i}\cos \varphi _{1}+ \vec{j} \cos \varphi_{2} + \vec{k}\cos \varphi
_{2}\right) ,  \label{01}
\end{equation}

\noindent where 
\begin{equation}
\varphi _{1}=k_{0}z-\omega _{0}t,\varphi _{2}=k_{0}x-\omega _{0}t,
\label{36}
\end{equation}

However, it can be shown that only one component of this force is reponsible for displaying large scale instability, which is the $(x,z)$ plan component. The force can then be reduced to:

\begin{equation}
\vec{F}_{0}=f_{0}\left( \vec{i}\cos \varphi _{1}+\vec{k}\cos \varphi
_{2}\right) ,  \label{01}
\end{equation}

Below are the expression of the Reynolds stresses as calculated in \cite{[1]}:

\begin{equation*}
T_{(1)}^{31}=\frac{D^{2}[2+Ra-2(1-W_{1})^{2}]}{\Xi _{(1)}},
\end{equation*}

\begin{equation*}
T_{(2)}^{31}=\frac{Ra[2+D^{2}-2(1-W_{2})^{2}]}{\Xi _{(2)}},
\end{equation*}

\begin{equation*}
T_{(1)}^{32}=\frac{-D^{3}[Ra+2[1-(1-W_{1})^{2}]]}{\Xi _{(1)}},
\end{equation*}

\begin{equation*}
T_{(2)}^{32}=\frac{-DRa[2+D^{2}-6(1-W_{2})^{2}]}{\Xi _{(2)}},
\end{equation*}

\noindent where

\begin{equation*}
  \Xi_{(1),(2)}=2[D^{4}+Ra^{2}+2D^{2}Ra+4[1+(1-W_{1,2})^{2}]^{2}+(2D^{2}+2Ra)[2-2(1-W_{1,2})^{2}]][1+(1-W_{1,2})^{2}].
\end{equation*}

\section{Large scale instability}

Let us write down in the explicit form the equations for nonlinear
instability:

\begin{equation}
\partial _{T}W_{1}+\nabla _{Z}T_{(1)}^{31}+\nabla _{Z}T_{(2)}^{31}=\Delta
W_{1},  \label{966}
\end{equation}

\begin{equation}
\partial _{T}W_{2}+\nabla _{Z}T_{(1)}^{32}+\nabla _{Z}T_{(2)}^{32}=\Delta
W_{2},  \label{970}
\end{equation}

\noindent where the components $T_{(1)}^{31}$, $T_{(2)}^{31}$, $T_{(1)}^{32}$
and $T_{(2)}^{32}$ of the Reynolds stress tensor are as defined in the
previous section.\newline

We shew in a previous paper \cite{[1]} that the linear instability can appear for specific values of $D$ and $Ra$. We
show below two figures representing the area (in grey) of the plane $(D,Ra)$
for which the discriminant is negative, this means that an instability can
appear. The first plot shows the conditions for a negative temperature
gradient and the second plot, for a positive one.

\begin{figure}[h]
\centerline{\includegraphics[width=0.35\textwidth]{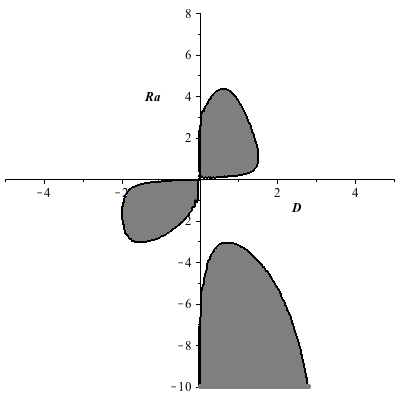}}
\caption{Instability condition with negative temperature gradient}
\end{figure}

\begin{figure}[h]
\centerline{\includegraphics[width=0.35\textwidth]{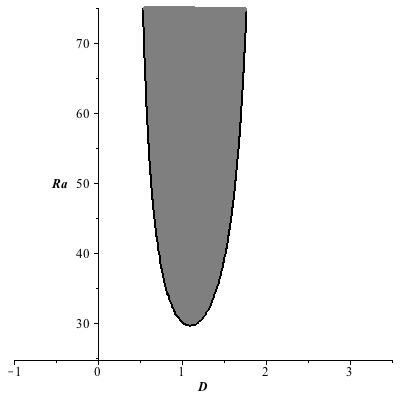}}
\caption{Instability condition with positive temperature gradient}
\end{figure}

\section{Instability saturation and non linear structures}

The increase of $W_1$ and $W_2$ leads to the saturation of the instabilty. As a result of the development and stabilization of the instability, non linear structures appear. The study of these stuctures is of interest. Then, we take back equations \eqref{966} and \eqref{970} allowing that $\partial_{T}W_{1}=\partial_{T}W_{2}=0$. Integrating these equations over $z$, we obtain the following system:

\begin{equation}
\nabla_{Z}W_{1}=T_{(1)}^{31}+T_{(2)}^{31}+C_{1},  \label{123}
\end{equation}

\begin{equation}
\nabla _{Z}W_{2}=T_{(1)}^{32}+T_{(2)}^{32}+C_{2}, \label{321}
\end{equation}

Where $C_{1}$ and $C_{2}$ are integration constants. Let introduce the new variables $X=1-W_{1}$, $P=1-W_{2}$. Then the equations \eqref{123} and \eqref{321} can be written:

\begin{equation}
\nabla_{Z}X=\frac{-D^{2}[Ra+2(1-X^{2})]}{\Sigma_{(1)}}-\frac{Ra(2+D^{2}-2P^{2})}{\Sigma_{(2)}}+C_{1},
\end{equation}

\begin{equation}
\nabla_{Z}P=\frac{D^{3}[Ra+2(1-X^{2})]}{\Sigma_{(1)}}+\frac{DRa(2+D^{2}-6P^{2})}{\Sigma_{(2)}}+C_{2},
\end{equation}

Where $\Sigma_{(1)}$ and $\Sigma_{(2)}$ are defined as below:

\begin{equation}
\Sigma_{(1)}= 2[D^{4}+Ra^{2}+2D^{2}Ra+4(1+X^{2})^{2}+(2D^{2}+2Ra)(2-2X^{2})](1+X^{2}).
\end{equation}

We have similar formulae for $\Sigma_{(2)}$ after having replaced $X$ with $P$.
Integrating this system using simple numerical tools allow us to display non linear structures. Below is the phase portrait of this integrated system of equations:

\begin{figure}[h]
\centering
\includegraphics[width=8cm]{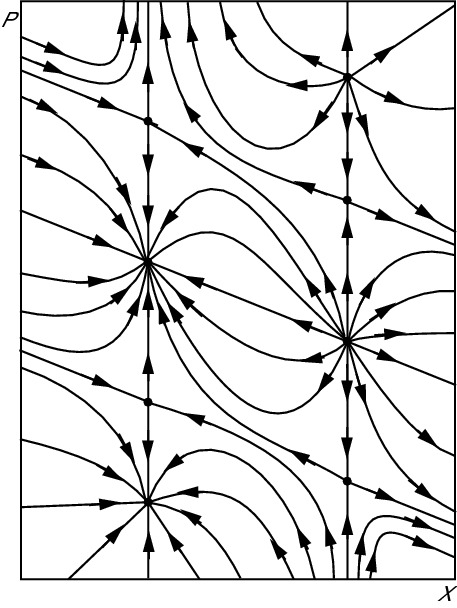}
\caption{Phase plane with stationnary points}\label{fig:somefiglabel}
\end{figure}

The phase portrait is unusual and doesn't contain any elliptic point, which means there is no periodical solutions. The only solutions we get which link different stationary points are helical kinks, namely, the velocity has a given constant value until a specific value of z, then it rotates helically and becomes constant again but with a different value of velocity. Below are some figures of these kinks:

\vspace{1cm}

\begin{figure}[!ht]
\centering
\includegraphics[width=5.4cm]{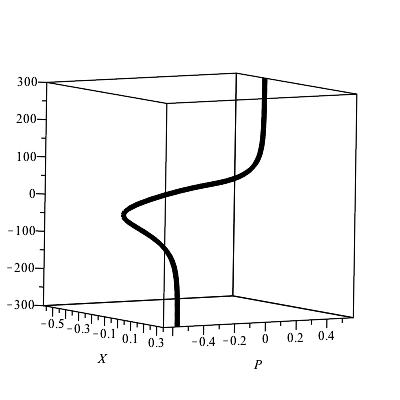}
\caption{Helical Kinks between two nodes, with $D=1$, $Ra=1$, $C_{1}=0,13$, $C_{2}=-0,1$.}\label{fig:somefiglabel}
\end{figure}

\begin{figure}[!ht]
\centering
\includegraphics[width=5.4cm]{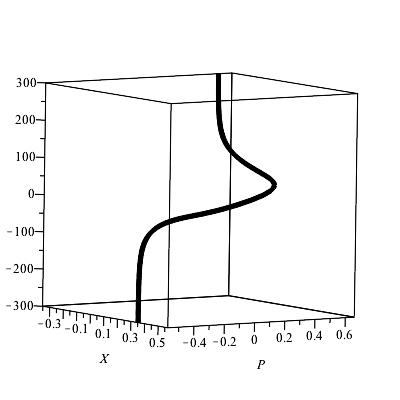}
\caption{Helical Kinks between a node and a hyperbolic point, with $D=1$, $Ra=1$, $C_{1}=0,13$, $C_{2}=-0,1$.}\label{fig:somefiglabel}
\end{figure}

\begin{figure}[!ht]
\centering
\includegraphics[width=5.4cm]{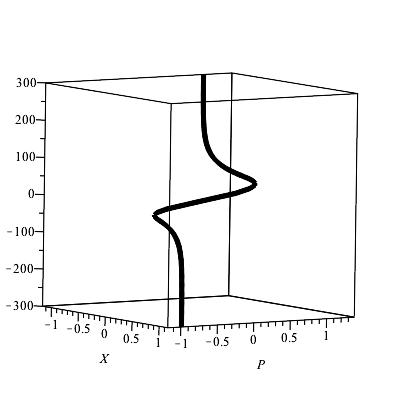}
\caption{Helical Kinks between two nodes, with $D=1,5$, $Ra=1,9$, $C_{1}=0,05$, $C_{2}=0,01$.}\label{fig:somefiglabel}
\end{figure}

\begin{figure}[!ht]
\centering
\includegraphics[width=5.4cm]{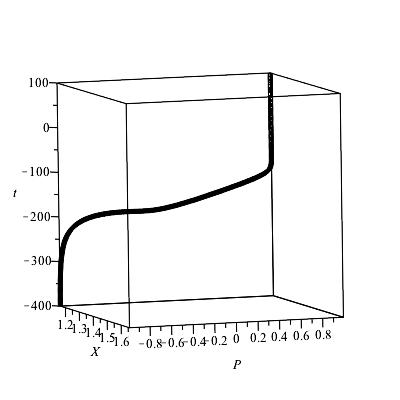}
\caption{Helical Kinks between a node and a hyperbolic point, with $D=1,5$, $Ra=1,9$, $C_{1}=0,05$, $C_{2}=0,01$.}\label{fig:somefiglabel}
\end{figure}

\newpage

\section{Conclusions and discussion of the results}

In this letter, we first briefly reviewed the main results of large scale instability in rotating and stratified flow found in \cite{[1]}, which was dedicated to the linear stage. Then we shew non linear large scale stuctures displayed by such a flow, which completes the previous study. These solutions show that in the non linear stage, the instability saturation leads to specific velocity profiles (helical kinks) for which the velocity tends to be constant for large values of $z$. These stuctures are of helical type and are the result of the saturation of the instability found in \cite{[1]}. Since the phase portrait doesn't contain any elliptic stationnary point there is no periodical solution but only rotational kink solutions.


\bigskip



\begin{thebibliography}{99}

\bibitem{[1]}  Tur, Chabane, Yanovsky, ``A new large scale instability in rotating stratified fluids driven by small scale forces'' Open Journal of Fluid Dynamics, 3, 340-351 (2013).

\bibitem{[14b]}  L. M. Smith, F. Waleffe, ``Generation of slow large scales
in forced rotating stratified turbulence,'' J. Fluid Mech. 451, 145 (2002).

\bibitem{[15]}  U. Frisch, Z. S. She, P. L. Sulem, ``Large-scale flow driven
by the anisotropic kinetic alpha effect,'' Physica D 28, 382 (1987).

\bibitem{[16]}  P. L. Sulem, Z. S. She, H. Scholl, U. Frisch, ``Generation
of large-scale structures in three-dimensional flow lacking
parity-invariance,'' J. Fluid Mech. 205, 341 (1989).


\bibitem{[18]}  S. S. Moiseev, P. B. Rutkevich, A. V. Tur, V. V. Yanovsky,
``Vortex dynamos in a helical turbulent convection,'' Sov. Phys. JETP 67,
294 (1988).


\bibitem{[21]}  G. V. Levina, S. S. Moiseev, P. B. Rutkevich, ``Hydrodynamic
alpha-effect in a convective system,'' Advances in Fluid Mechanics 25, 111
(2000).


\bibitem{[26]}  A. V. Tur, V. V. Yanovsky, ``Non linear vortex structures in
stratified fluid driven by small-scale helical force'', Open Journal of
Fluid Dynamics 3, 64 (2013).

\bibitem{[27]}  R.Marino, P.D.Mininni, D.Rosenberg, and A.Pouquet,
``Emergence of helicity in rotating stratified turbulence'',
Phys.Rev.E87,033016 (2013).
\end{thebibliography}
\end{document}